# Light Efficient Flutter Shutter

Moshe Ben-Ezra
August 2012

moshe@ben-ezra.org

## Abstract

Flutter shutter is technique in which the exposure is chopped into segments and light is only integrated part of the time. By carefully selecting the chopping sequence it is possible to better condition the data for reconstruction problems such as motion deblurring, focal sweeping, and compressed sensing. The partial exposure trades better conditioning for less energy. In problems such as motion deblurring the available energy is what caused the problem in the first place (as strong illumination allows short exposure thus eliminates motion blur). It is still beneficial because the benefit from the better conditioning outweighs the cost in energy.

This documents is focused on light efficient flutter shutter that provides better conditioning and better energy utilization than conventional flutter shutter.

The document is related to the following patent and patent applications:
1. US Patent 7,756,407 Method and apparatus for deblurring images.
2. US Patent App. 2011.024.2334 Time Interleaved Exposures and Multiplexed Illumination.

# 1. The problem

In order to obtain an image with sufficient SNR, it is necessary to integrate the signal (image) over a finite time interval t > 0. If during this time the signal is changed, for example due to motion, resulting with motion blur then the signal restoration is dependent on the signal itself, and on the windowing function – in our case the exposure pattern.

In this document we refer only to binary on/off temporal windowing function, and for comparison reasons we will use 52 bins (the signal shall be 64 bins for use with FFT, but the last 12 will always be windowed-out and have a value of 0).

A conventional exposure has a Rect temporal windowing function. Figure-1 shows the power spectrum in dB of the Rect windowing function. As can be seen in the graph, the power spectrum falls sharply (the graph is logarithmic) at multiple points. The frequencies corresponding to these points are practically lost (due to noise) which makes the recovery difficult.

**Sequence of 52 chops:**
1111111111111111111111111111111111111111111111111111

Figure 1 - Rect power spectrum in dB, X=relative frequency

**Note: 22-Aug-2015**

**Note that in order to have same \*energy\* this sequence length should be half the length of the flutter shutter sequences because of duty cycle differences.**

## 2. Flutter shutter solution and limitations

To address this problem, Raskar proposes using coded exposure to better condition the problem. A pattern such as the MURA pattern can be used, but because the MURA pattern is not optimal for zero padded sequences, Raskar proposed a different sequence found by exhaustive search. Figure-2 compares the power spectrum of the sequence proposed by Raskar to the power spectrum of the Rect shown above. Improvement of 5-20 db or more is achieved in most frequencies.

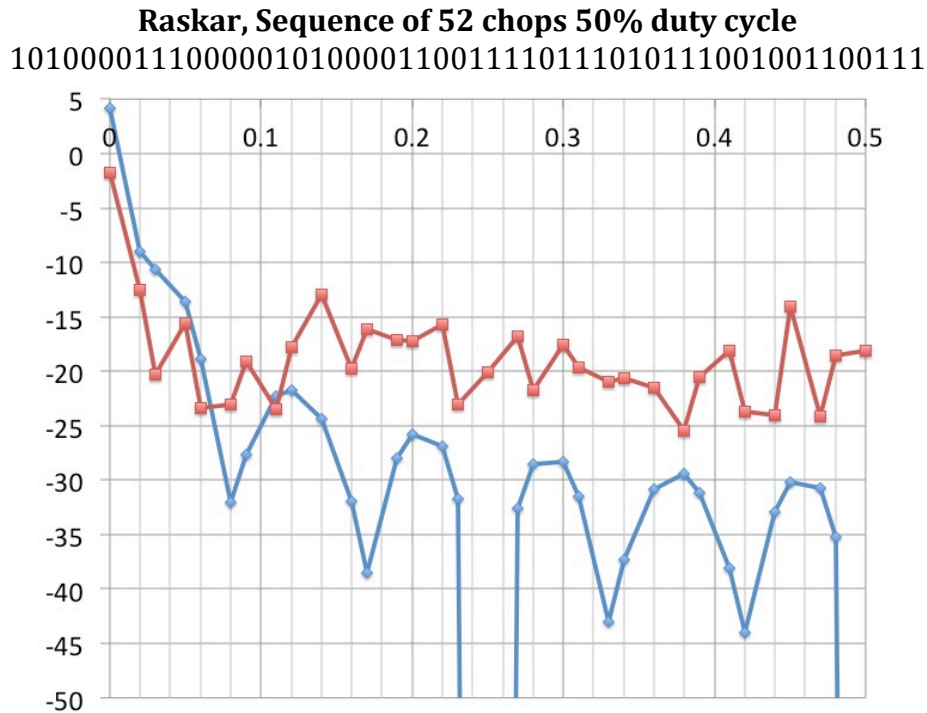

Figure 2 - Power spectrum of Raskar's sequence compared to Rect's

However, the duty cycle of the pattern above is only 50% resulting with a lost of half the available light. Furthermore, if the shutter is implemented using a ferroelectric device, that uses light polarization for shuttering, it will result with additional 50% (or more) lost of light, which puts the overall efficiency of the system at 25% or two stops less to start with.

## 3. Exposure pairs

The main observation for this solution is that once an optimized sequence is found, the power spectrum of the sequence is the same as the power spectrum of its compliment (up do DC value if not 50% duty cycle). In other words, flipping 0 and 1 in the sequence will flip the *phase* in the Fourier domain by 180 degrees, but will not change the *length* of the vectors.

Therefore, if we can collect the light form the sequence and its compliment, both will be well conditioned, and the total duty cycle will be close to 100%. The simplest (but not limited to) way to so this is to process each part separately and then add the resulting images together.

Note that while FFT is a linear operator the deblurring operator is not linear, and therefore signal to noise performance is not straight forward computed and will depend on the specific operator used.

Also note that the complement is taken over the windowing function not the signal as the signal may vary.

### 3.1 Optomechanical implementations

Figure-3 shows two mirror-based configurations that can implement the window-complement pairs. Figure-3(left) shows a tilt-switch mirror that can send the light to one sensor or the other according to the coded sequence. Such setup can be implemented using a DMD device; in this case spatial windowing is also possible. Figure-3(left) shows a rotated mirror wheel where transparent coated parts hard-code the sequence. Light is transmitted or reflected according to the coded coating.

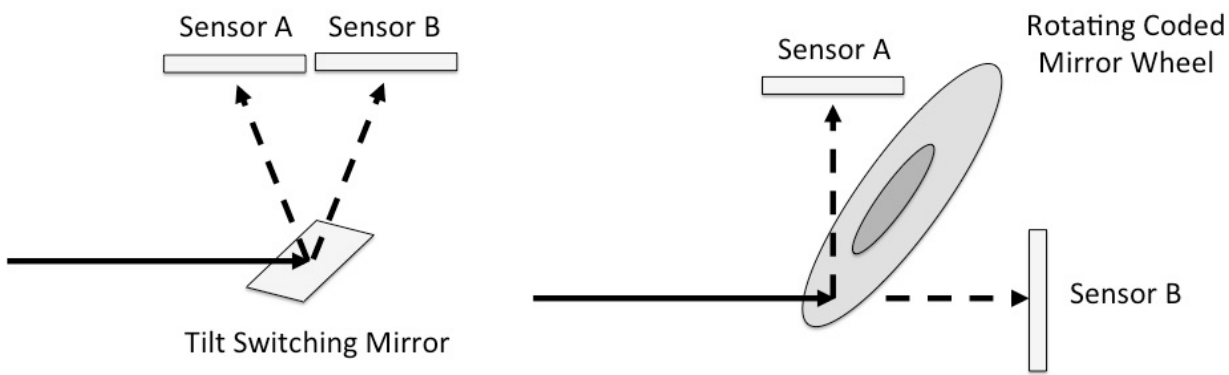

Figure 3 - Mirror-based configuration

## 3.2 Electronic CCD implementations

The optomechanical configurations are relatively big and complex; and electronic implementation would be preferred.

Figure-4 shows an implementation of a chopped binary windowing function and its complement using an interline CCD.  Interline CCDs have half their pixels shielded from light.  These pixels are used as electronic shutter and also allow fast capture one frame while reading the other.  An interline CCD can be clocked to transfer the charge "horizontally", thus saving the latest captured image at the light shielded pixel, and then these pixel can be clocked to transfer the charge "vertically" for reading.  In principle the CCD can be clocked to transfer the charge "horizontally" in both directions thus alternating the exposure between two adjacent columns, one is a complement of the other. This capability is probably used to implement external trigger mode 4,5 in some of Pointgrey's CCD cameras.  However, in modes 4,5 only one image can be read, see note below.

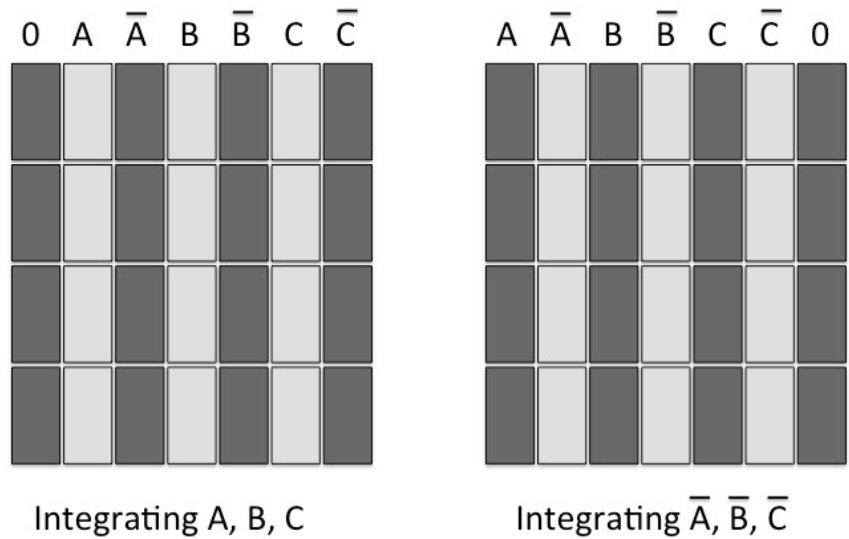

Figure 4 - Window and Complement implementation using interline CCD

Same principle applies for frame-transfer CCDs as well.  CMOS sensors cannot move change the way CCD sensor can. However, CMOS sensors can use switch in the imaging elements as described in US patent application: US20110242334 [2].

**Note:** To avoid image smearing during readout, both interline CCD and frame transfer CCD require a physical shutter to block the light during readout. For this reason mode 4,5 mentioned above cannot provide both window and its complement data. To overcome this limitation see next section.

### 3.3 Electronic CCD implementations

To overcome the need for physical shutter in CCD sensors, two configurations are presented – modified interline CCD and hybrid interline/frame-transfer CCD.

Figure-5 shows a modifies interline CCD. In this configuration an additional shielded column is added for each pixel. This column is used during readout to protect both the windowed signal and is complement from smearing.

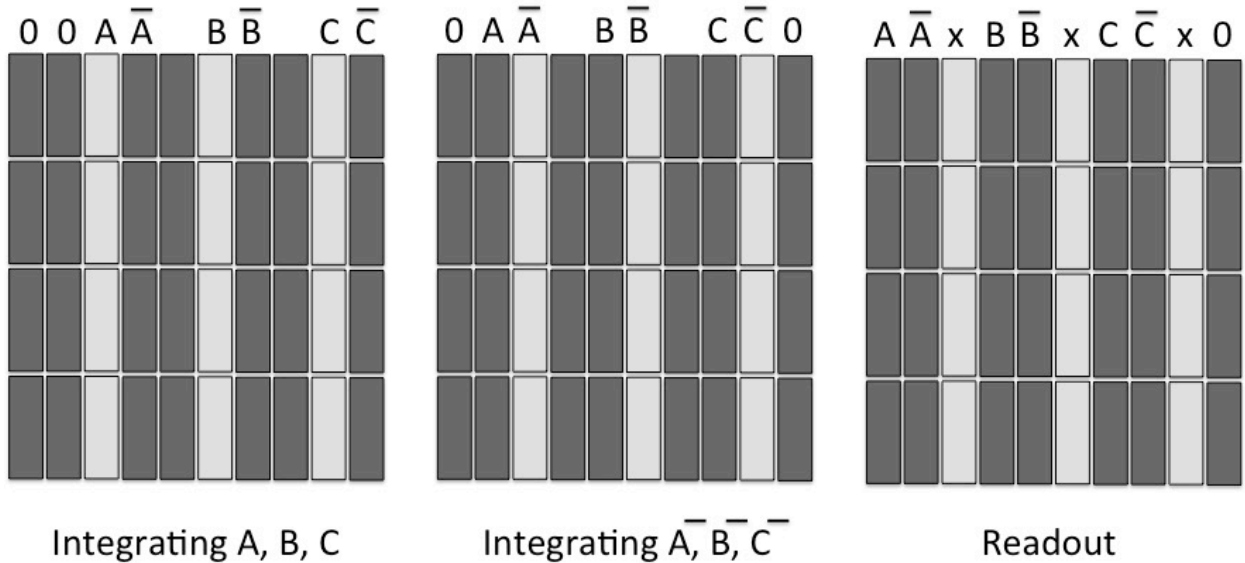

Figure 5 - modified interline CCD

Figure-6 shows a hybrid interline frame transfer CCD. In this configuration an additional shielded area of the same size is. The entire frame is transfer to the shielded area and then read column by column. This configuration only require horizontal transfer. Other configuration can combine horizontal and vertical transfer, for example by placing the shielded area below the interline area. Additionally, If the two parts can be clocked separate, it is possible to capture two frame (window and its complement) while the previous two are being read enabling fast frame rate.

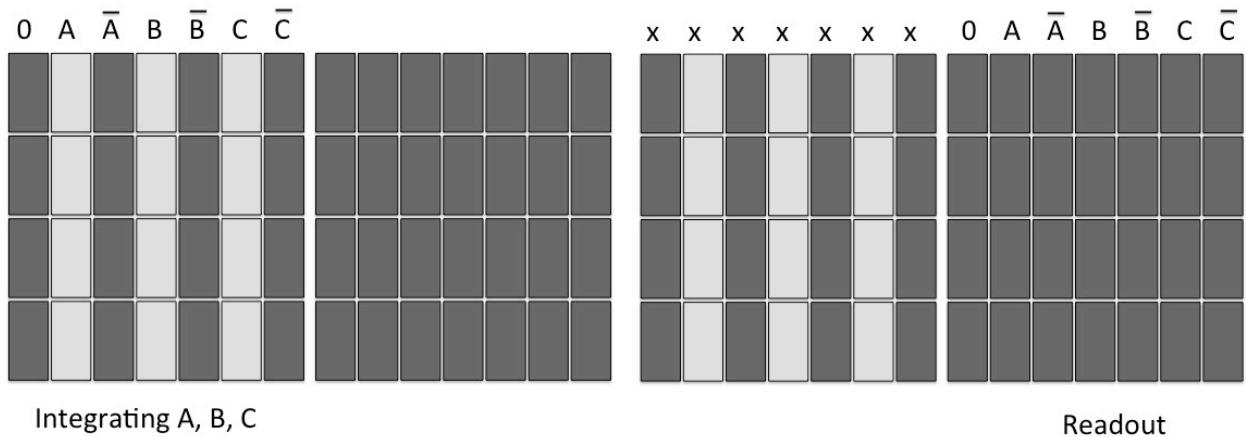

Figure 6 - Hybrid interline frame transfer CCD

## 4. One of N (small N)

When using window / compliment pair, the frequency attenuation of the window is determined by the first sequence (the complement is the same) and there for it is important to find the best sequence possible. As there are N choose K combination this is a difficult task even if we search only for one value of K (= n/2) for the 52 elements mentioned here it will take 15 years to search all combination in a rate of one million a second.

However, instead of finding one best sequence, we can find 3 or more sequences such that at each location one and only one of the sequences will have a value of one and the rest will have a value of zero. We can also request that each sequence will have approximately 1/N values that are one – but this is not mandatory.

The first thing that this method does is increasing the search space even more. However, it also provides an additional degree of freedom. Now each sequence can have its own spectral attenuation. If we look at the combined information (for example selecting the good frequencies form each image produced by the different windows, or weight them accordingly) we can see that it becomes relatively easy to find N good sequences. For example, the three in Figure-7 were selected using best of 100 random samples – no optimization (a genetic algorithm can be very effective for this type of search) was done.

**Best one of 100 random samples**
1011001100000100000110100000000101110001110010010100
0100000011110000010000010001100010000000001001000001
0000110000001011101001001110011000001110000100101010

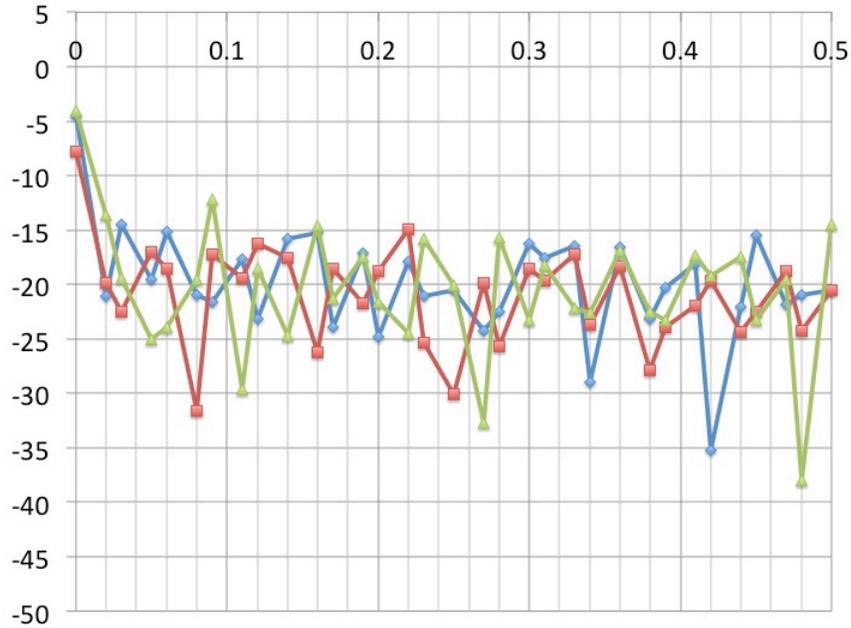

**Figure 7 - Power spectrum of 3 sequences**

We can see that none of the three sequences is optimal and that the power spectrum of the sequences is generally not correlated. If we select the best frequency band from each window and combine them, we can see in Figure-8 that the combined response is as good or better as the optimized sequence found by Raskar.

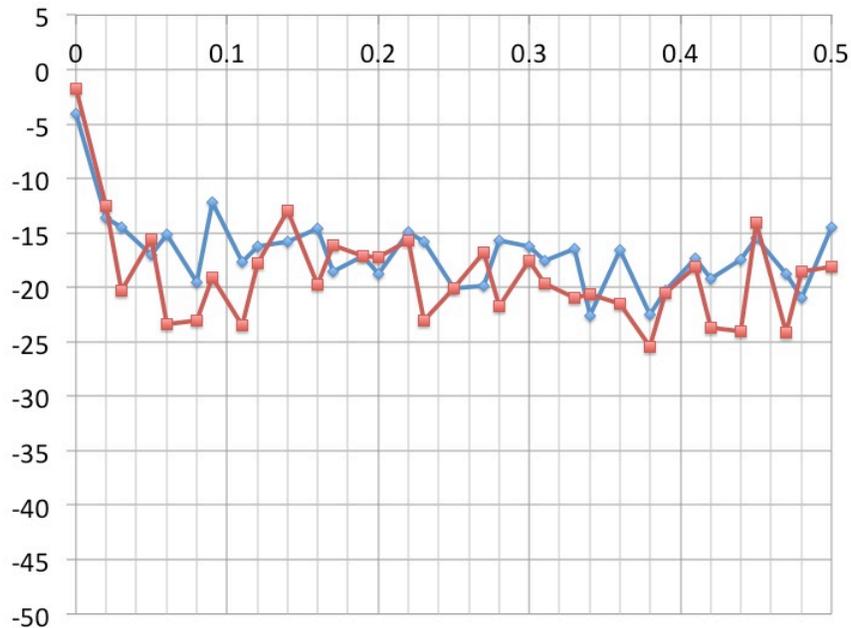

**Figure 8 - Combined window response (blue), Raskar's window response (red)**

The implementation of the 3-windows CCD (or CMOS) is similar to the implementation of the 2-windows mentioned above. Figure-9 illustrates interline CCD implementation for 3-windows with physical shutter. Electronic shutter can be added in similar way to the 2-windows implementation though, a hybrid interline/frame-transfer would be recommended for density considerations.

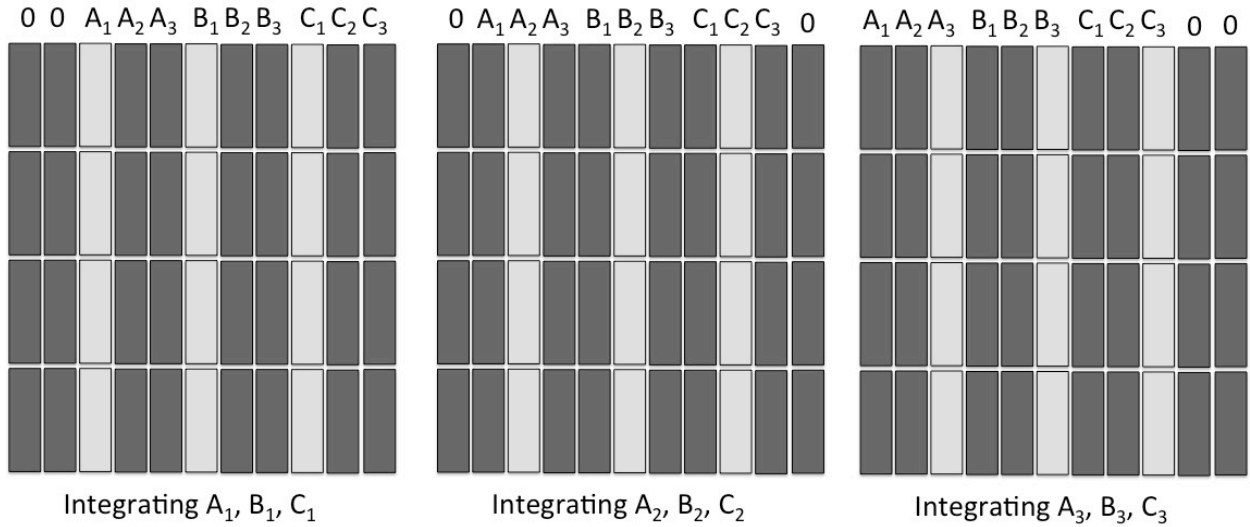

**Figure 9 - modified interline CCD for 3-windows implementation**

## 5. Optimization using Genetic algorithm (1 of 3 codes)

Flutter shutter codes, including one-of-N can be coded as string of numbers for example the string:

"221214411124122441241121122124144112124441212442242"

codes the following 3 sequences, in which only one bit is on at each time interval.

"000001100001000110010000000010111000001110000110010"
"1101000000100110001000100110100000001010000101001101"
"001010011001000010011011001001000110100001010000000"

Such coding can optimized by a standard genetic algorithm framework, for example the results below used an algorithm with population: 1000, selection which is proportional to the merit (fitness) function value, probability of 0.9 for crossover, and 0.01 for mutation. I also used three elite member to keep the best values (so far) intact.

The main difference between different experiments rises from merit functions used. Three different merit functions were tested and the results are described below:

## 5.1 Max Min of Spectral Power merit function

In this merit function we compute the power spectrum distribution of the each of the three sequences. For each sequence we take the minimal (worse) value and the return the maximum of these values as the merit function. In principle it is sufficient, but not necessary to have one good sequence to maximize the merit function. As it turned out the algorithm actually produces results that had one sequence significantly better than the other two. The max of three improved the result even more but not dramatically so. In fact the best of three sequences was even better than the sequence found by Raskar et al. Because only few examples were tested it is not clear if this behavior it typical or just received by chance.

The best result obtained for this experiment was (green is the best sequence):

Merit: -21.4382
22124414141124124412241112412442442422124422114 44144

S1: 00001101010001001100010000100110110100001100001 11011
S2: 11010000000100100011000010010010010110100110000 0000
S3: 00100010101100100010001110010000000001000001100 0100

Note that the sequence and it compliment (inverted) sequence can both be used and will have the same spectral power distribution. The spectra power distribution and comparisons are given at the following charts.

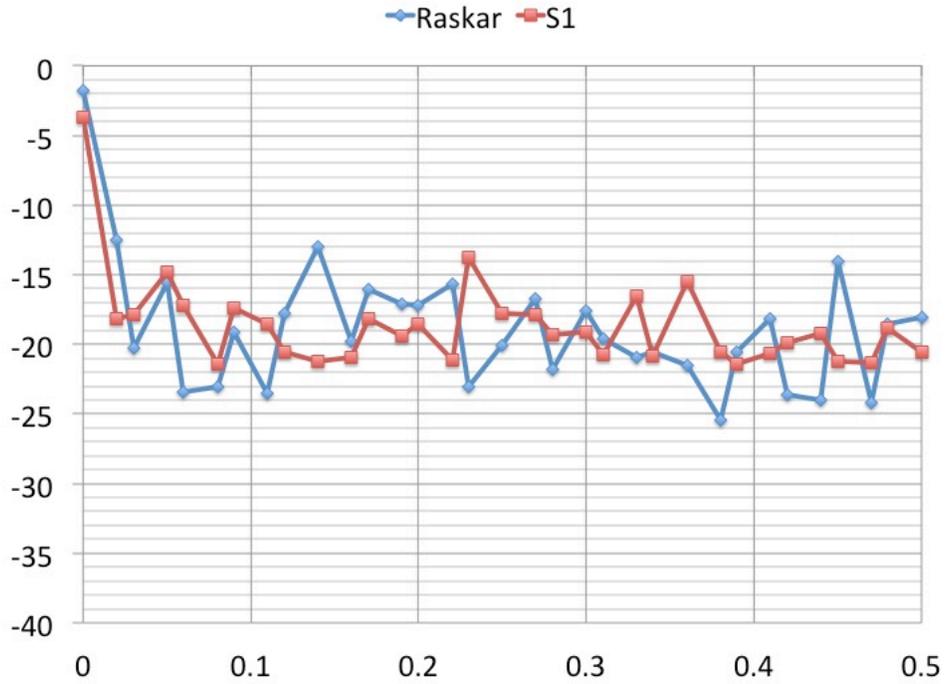

**Figure 10 - Best of three compared to Raskar's sequence**

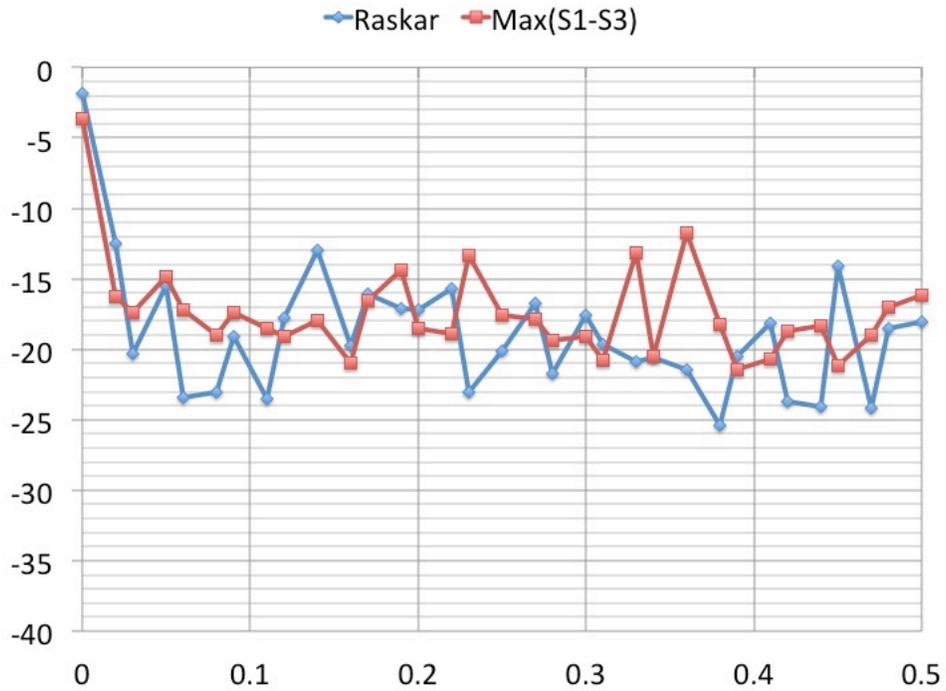

**Figure 11 - Max od (s1.. S3) compared to Raskar's sequence**

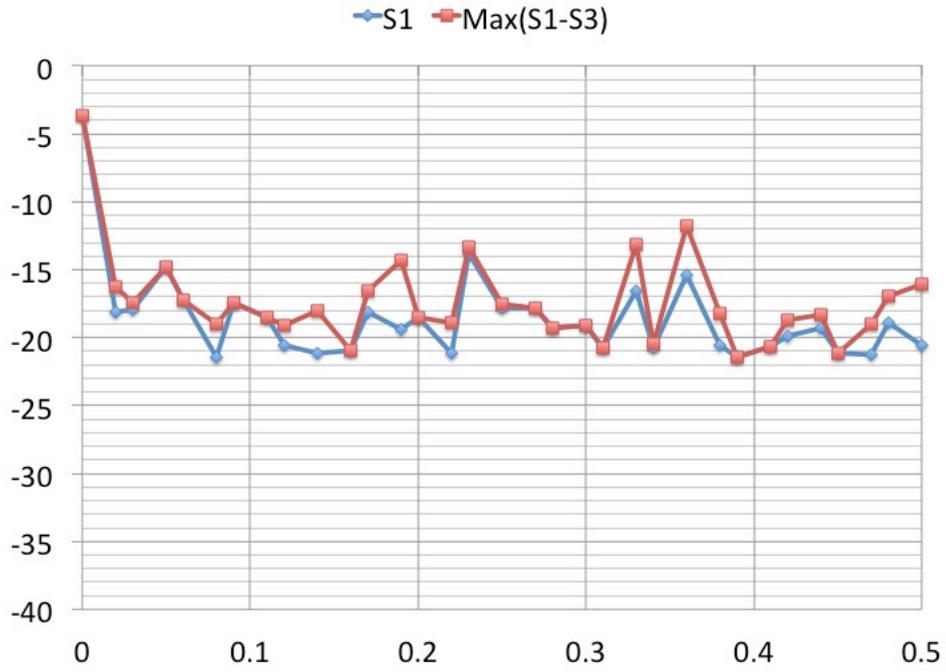

**Figure 12 - S1 Compared to Max(S1-S3) difference is not dramatic**

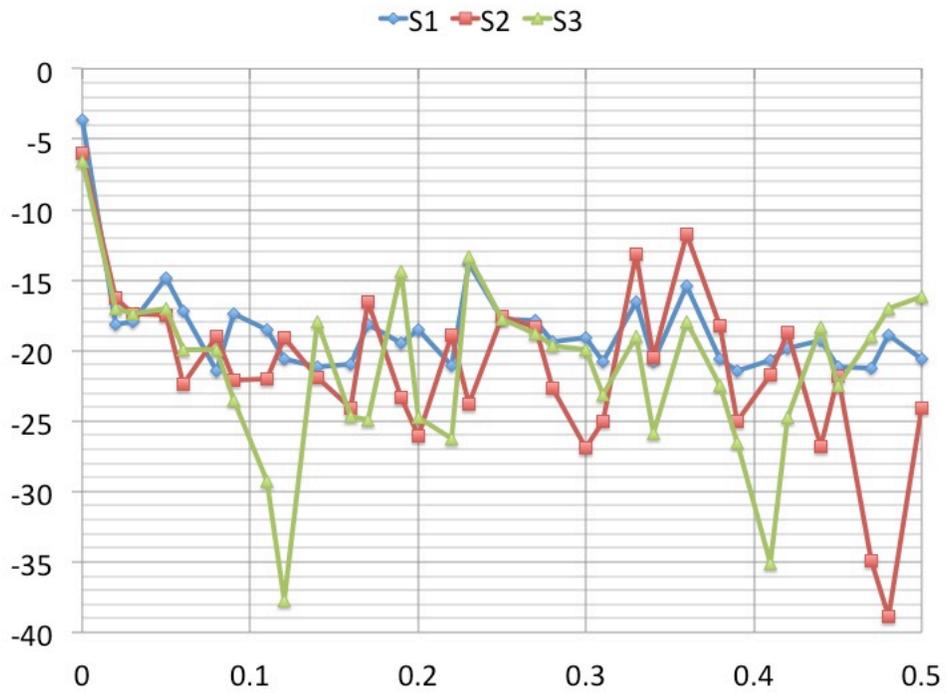

**Figure 13 - All 3 sequences. Each contributes to the max, though S1 is best**

Similar results were obtained using the sequences:

Merit: -21.5469

1212222141422244221122444224241121424422424241241242

S1 : 00000000101000110000011100101000101100101010010010
S2 : 01011110000111001100110001101000100100110101001001 01
S3 : 10100001010000000011000000000011010000000001001000

And:

Merit: -21.7312

2211424222421411121144411124121212214214141122112142

S1: 00001010001001000000111000010000000010010100000000 10
S2: 11000101110100000100000000100101011001000000110010 01
S3: 00110000000010111011000111001010100100101011001101 00

Same graphs as before for these sequences are shown below:

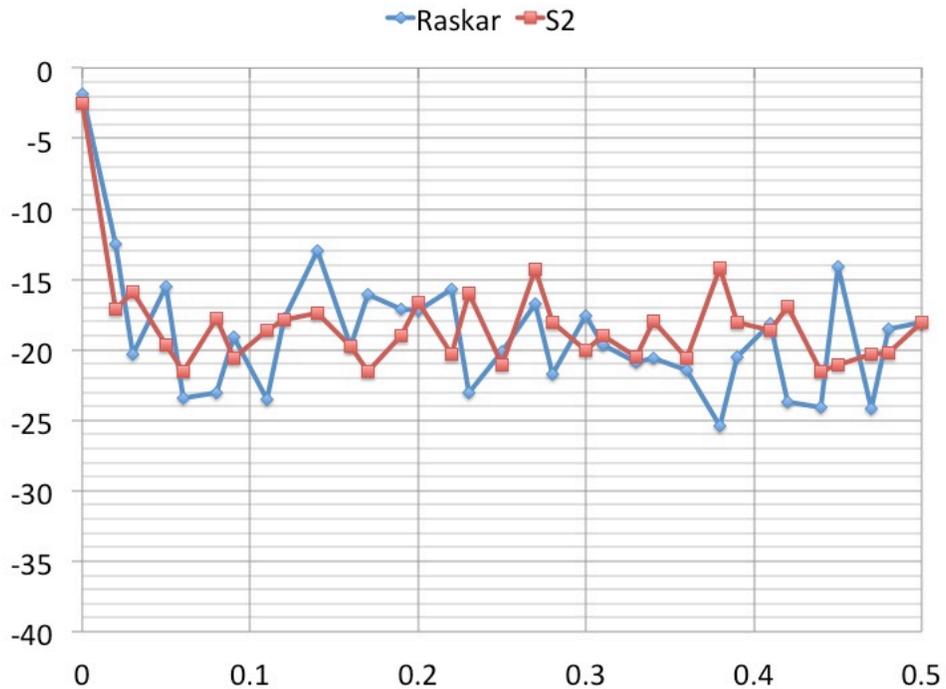

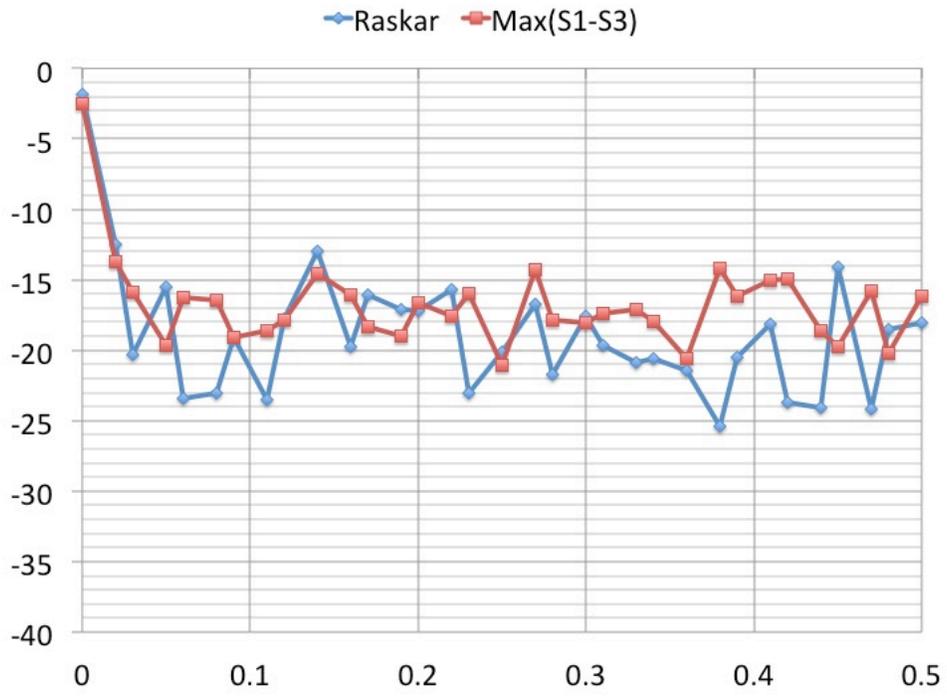

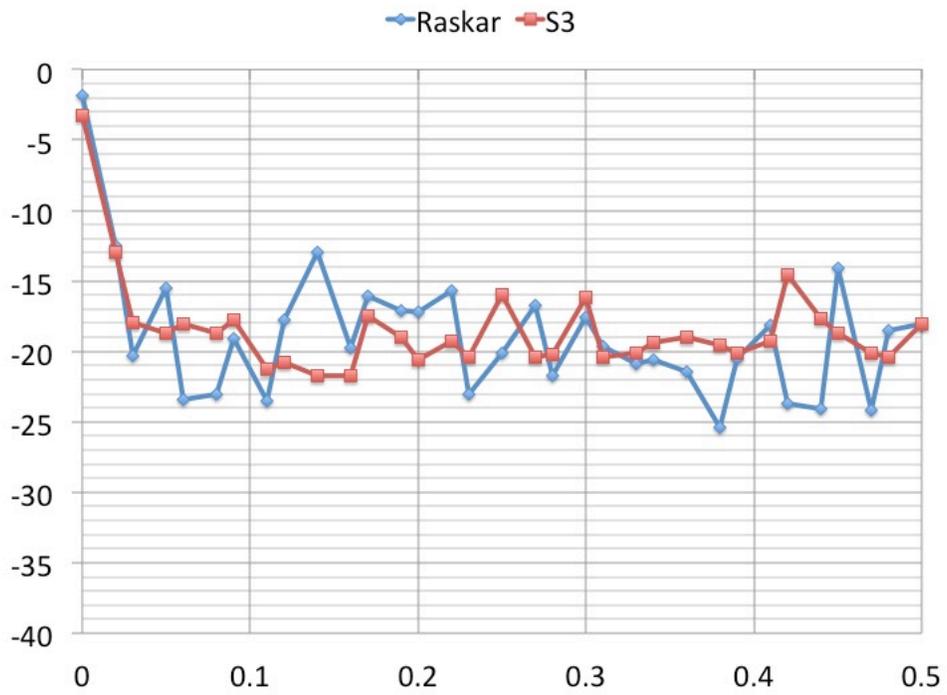

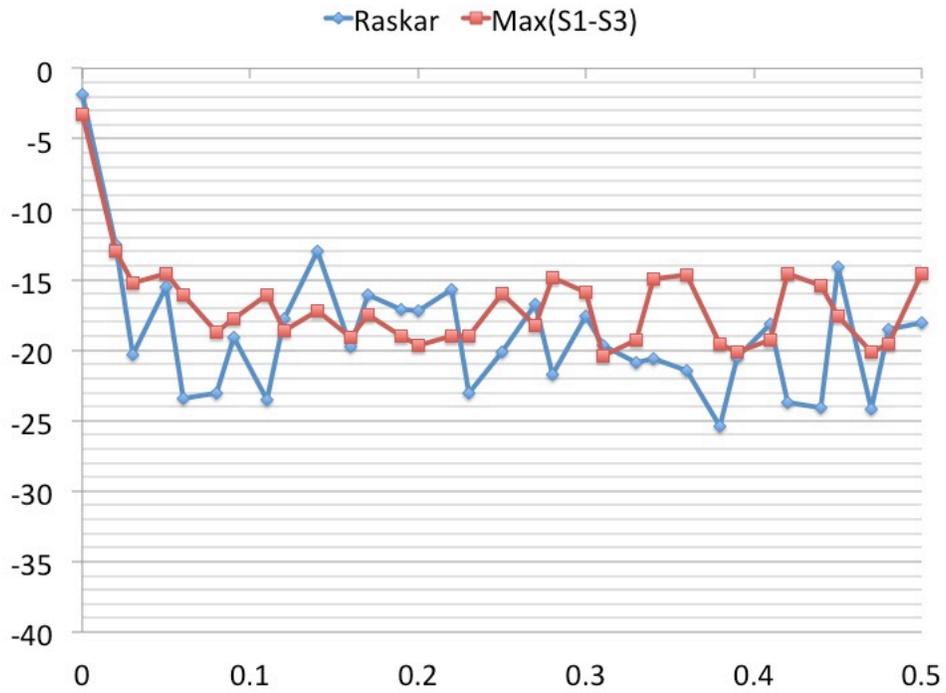

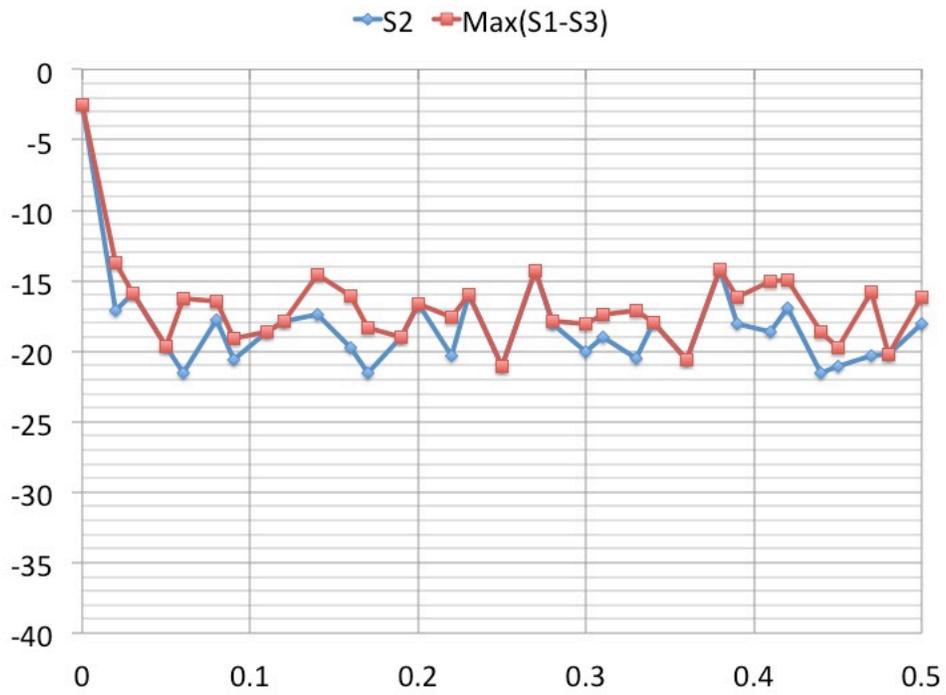

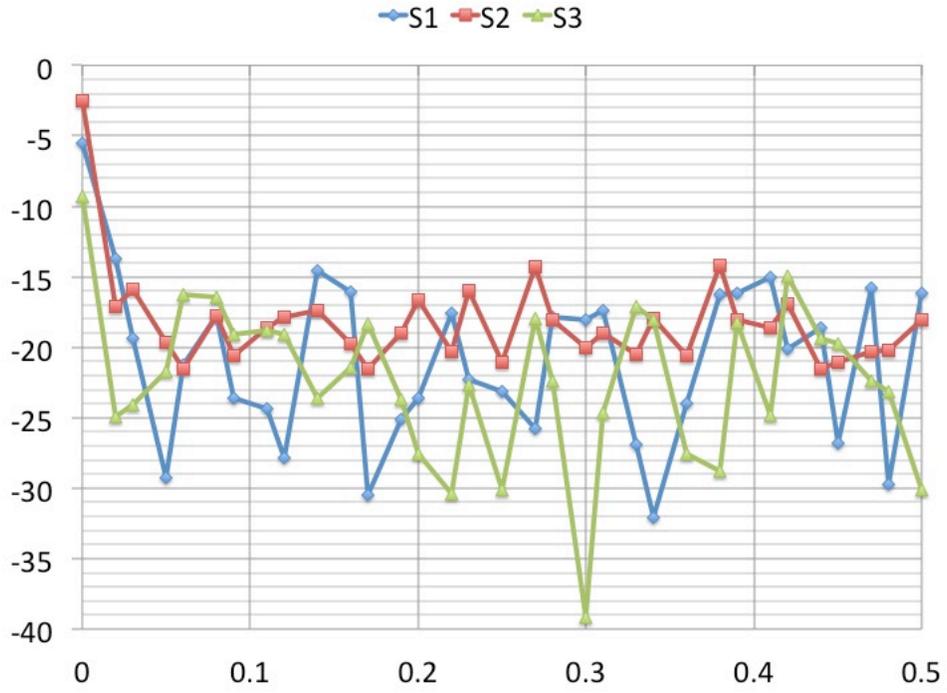

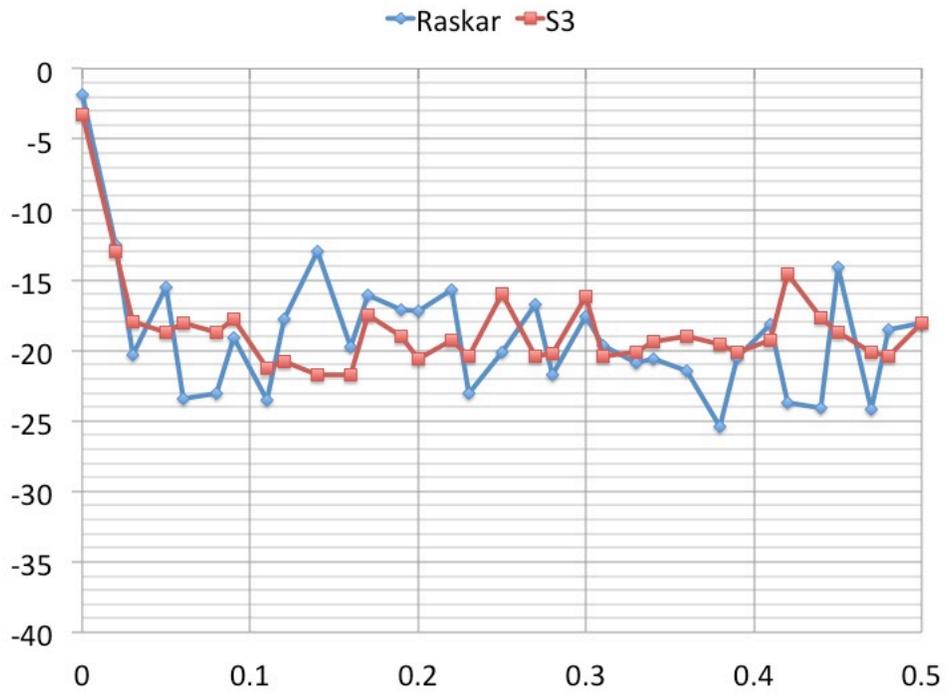

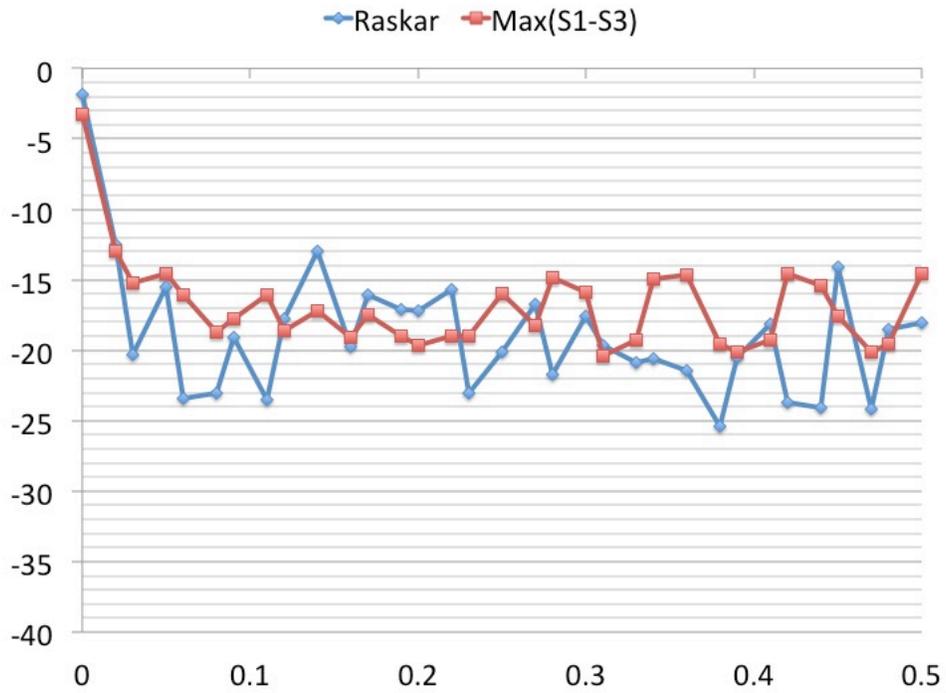

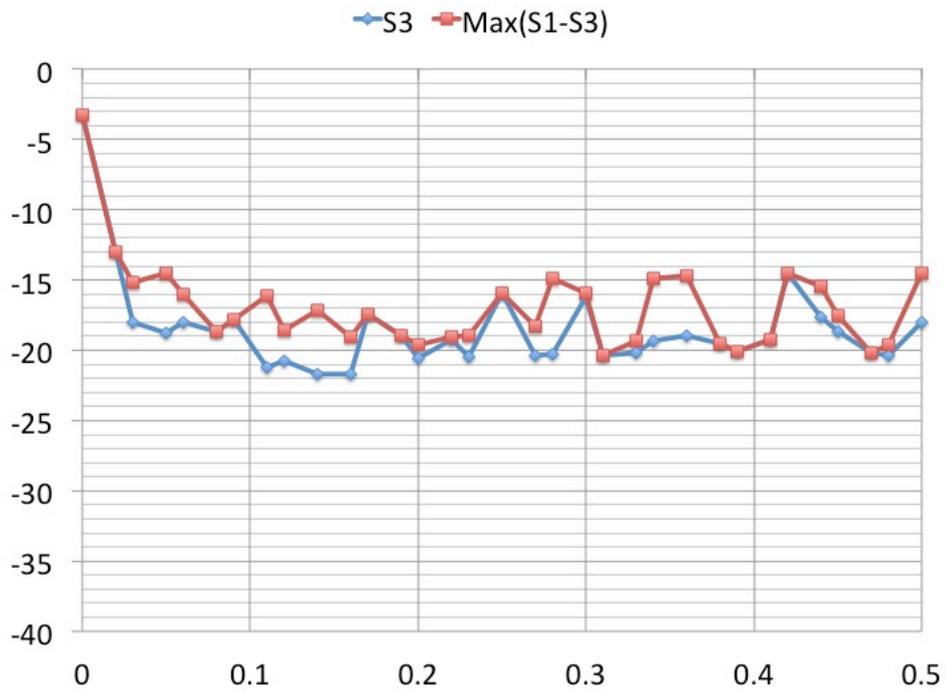

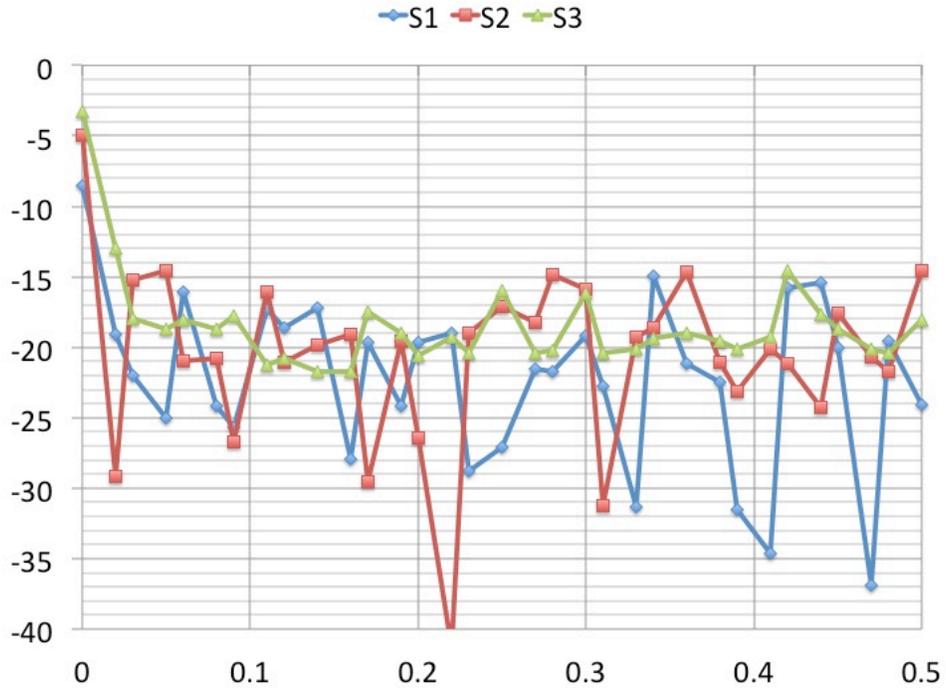

## 5.1 Average Min of Spectral Power merit function

This merit function still takes the lower bound (min) of the power spectrum of each sequence, but then averages them. This favors more uniform distribution of the three sequences – but also reduces the best result. The sequences and charts for this merit function are shown below.

Merit: -25.6353

221214411124122441241121122124144411212444 1212442242

00000110000100011001000000000101110000 1110000110010
110100000010011000100010011010000000 1010000101001101
00101001110010000100110110010010001101 00001010000000

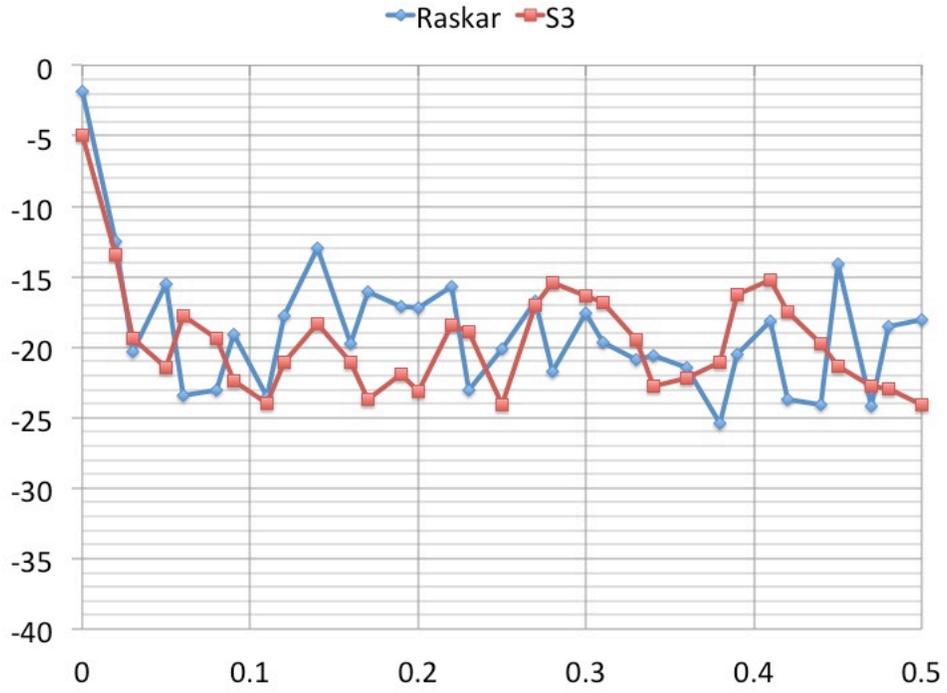

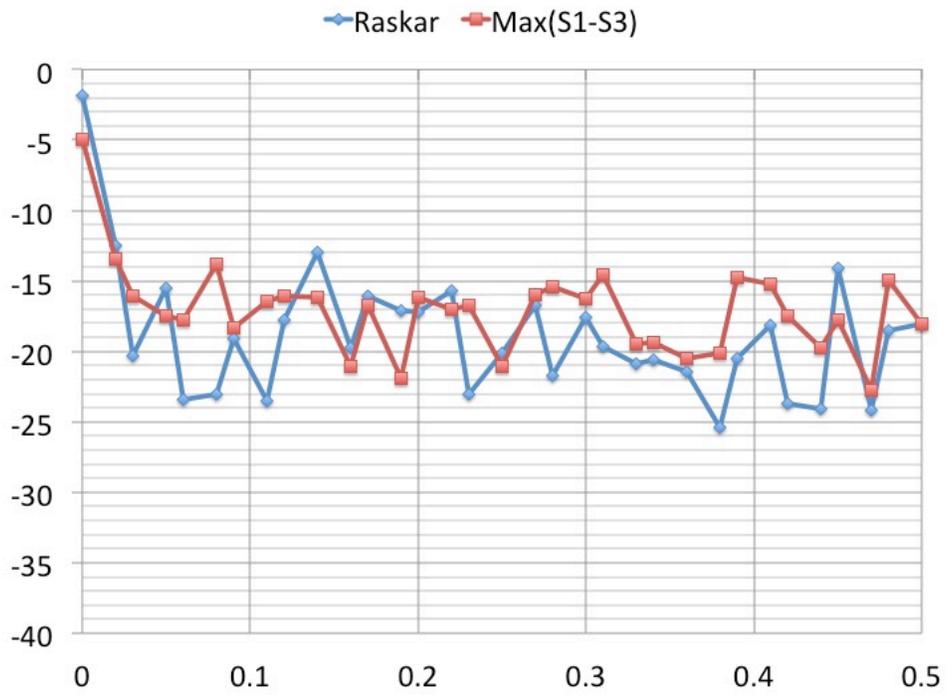

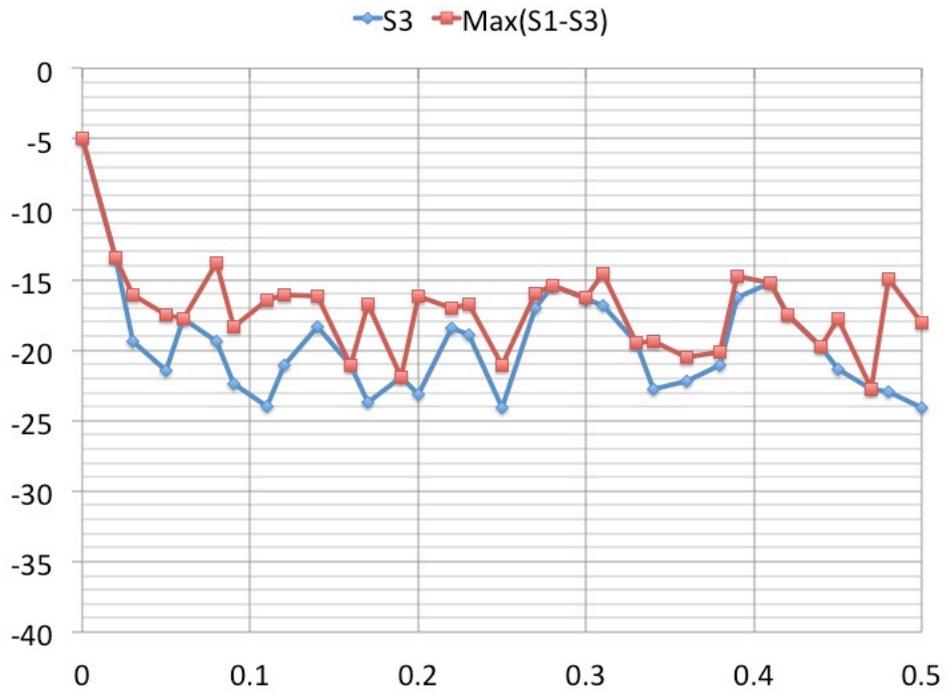

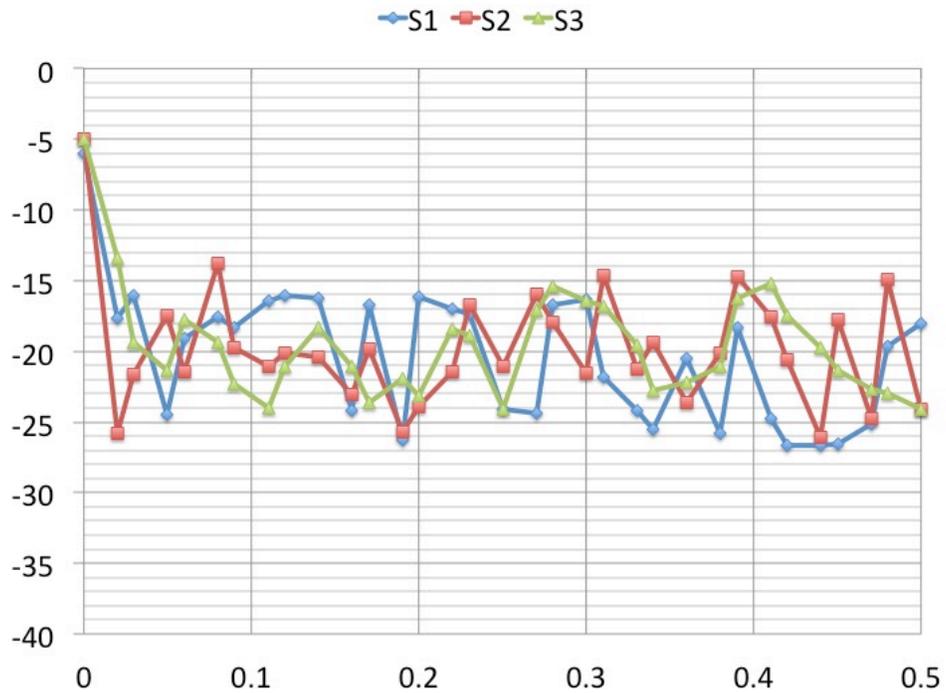

## 5.1 Average Pairs of Max Min Spectral Power merit function

When one sequence is determined, there is still some freedom in setting the other two.
This merit function first finds the min (worst) value of each sequence, then find the max of pairs (S1,S2, S1, S3) and average them. This results with an in-between merit function two sequences are optimized, as shown in the results below.

Score: -21.6372

22214211244212221111221421412112441212422411142211211

00001000011000000000001001000001100001001000100000
11100100100101110000110010001001000101011000001101100
000100110000100011110010010101100010100000111001011

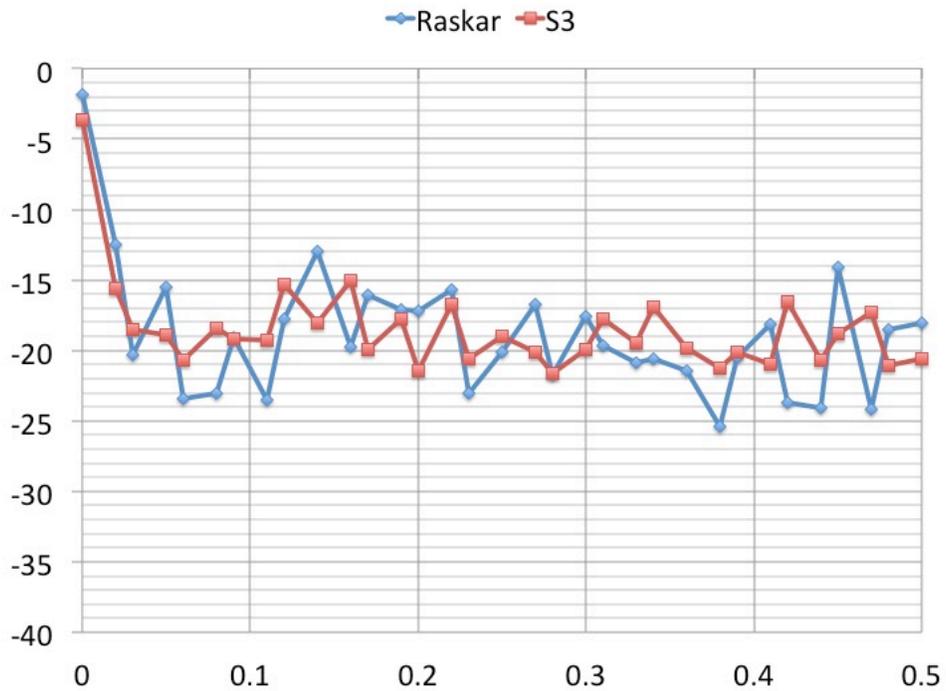

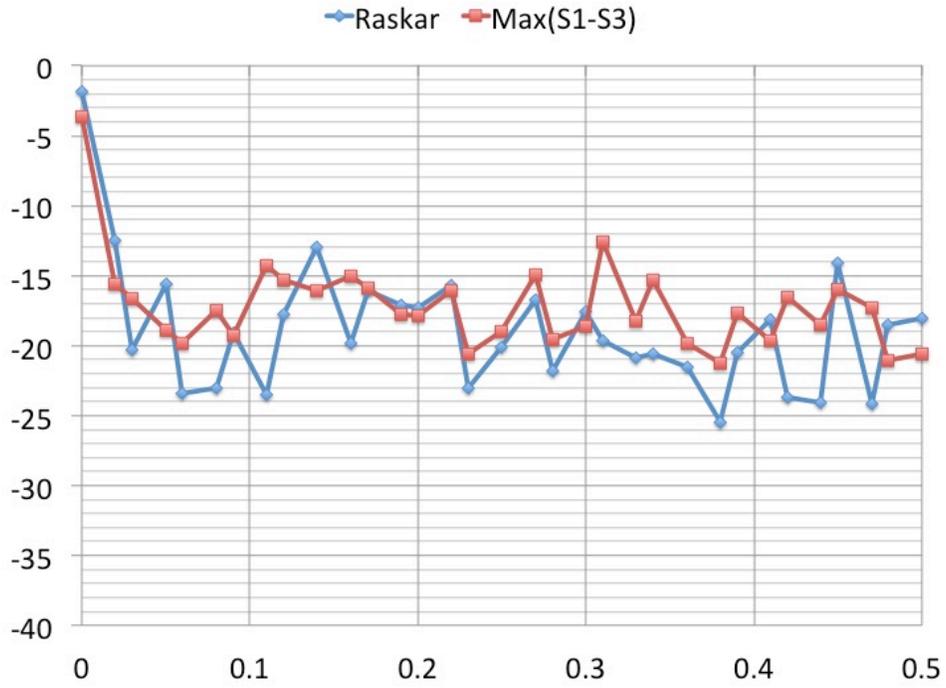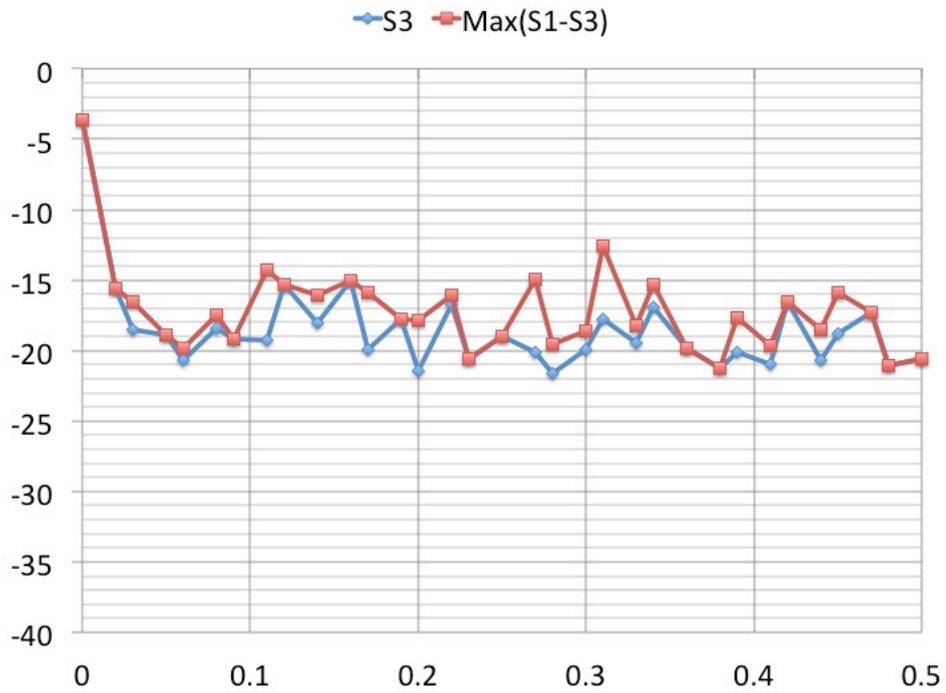

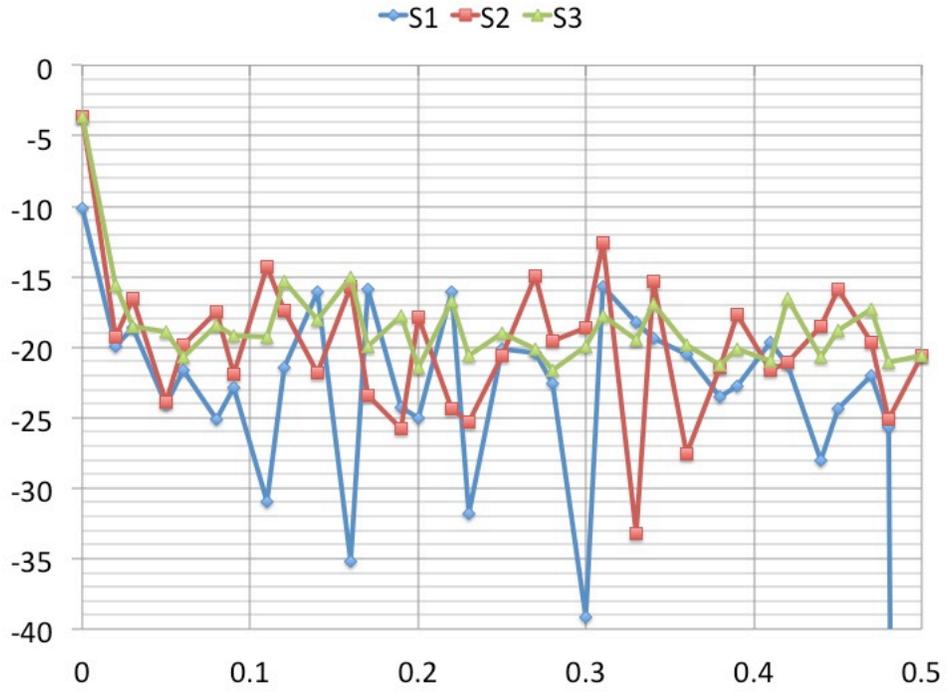

## 6. Image testing simulation results

This section describes simulation results using the image shown in Figure-14.  For each test the image was blurred using different kernels of size 52 pixels (for comparisons).  Gaussian noise with mean = std = alpha * 0.01 was added to the image, where alpha is proportional to the exposure of each test.   Brightness was the adjusted to same value  (gain). Results are shown below:

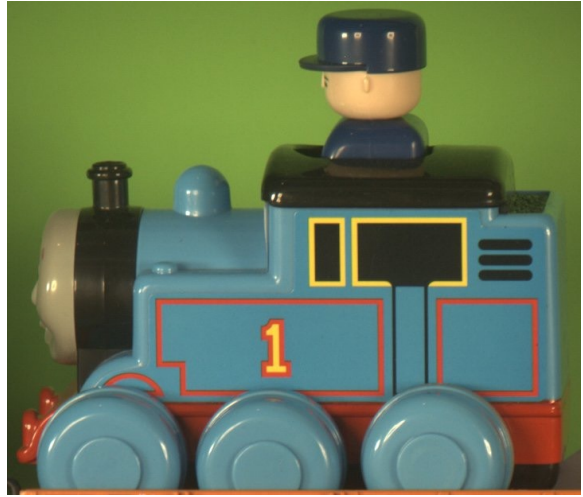

Figure 14 - Ground truth image

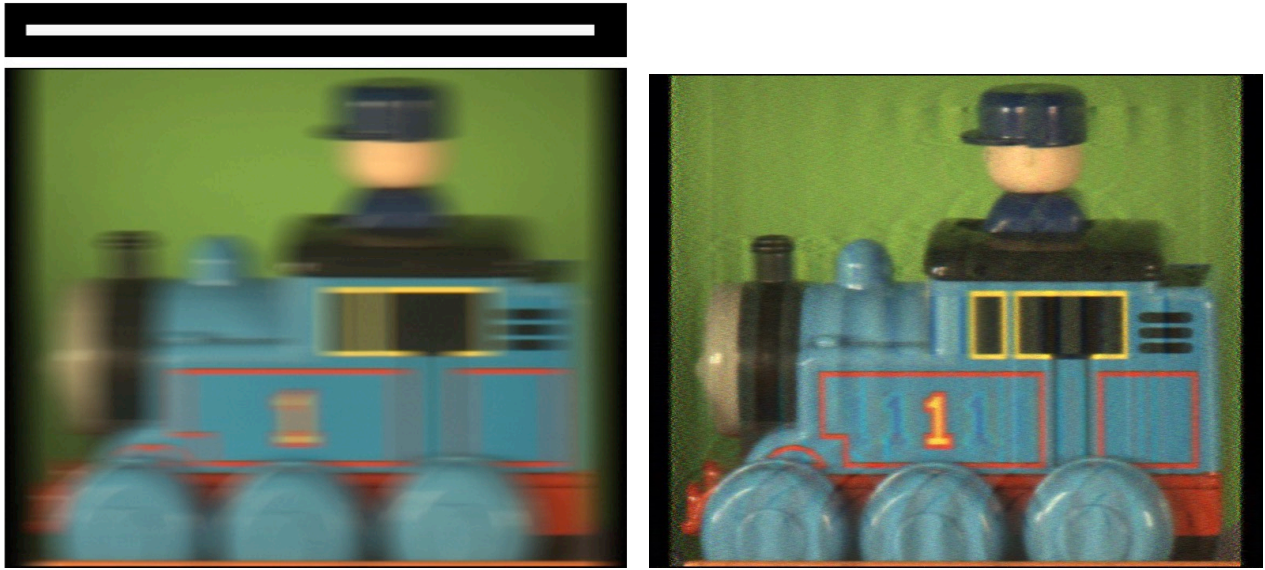

Figure 15 - Flat 52 pixels PSF blurred image  and deblurring result

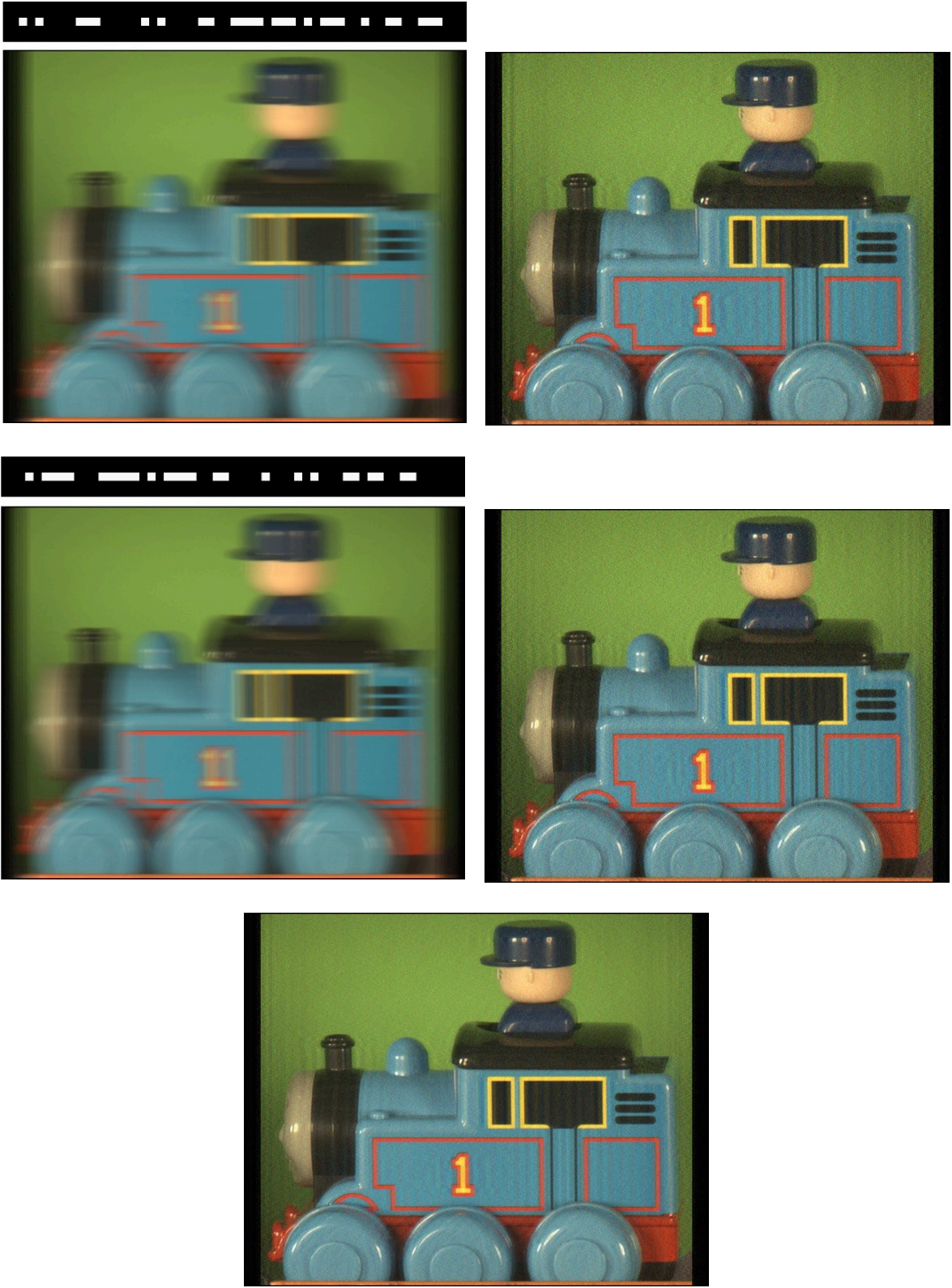

**Figure 16 – Top: Original Raskar's kernel blurred image and deblurred result. Middle: Inverted Raskar's kernel blurred image and deblurred result. Bottom: average of the two results (noise is uncorrelated).**

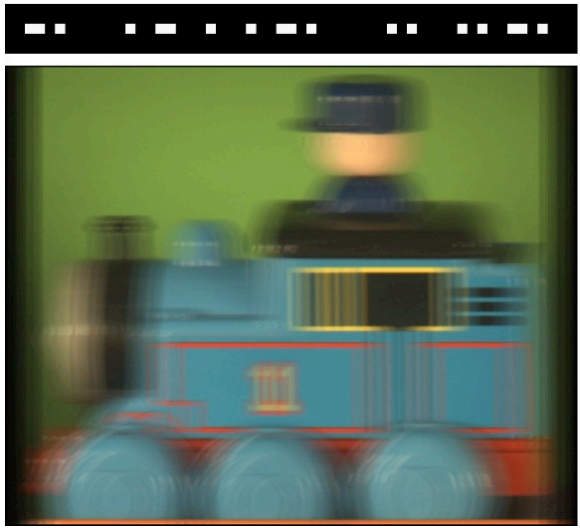 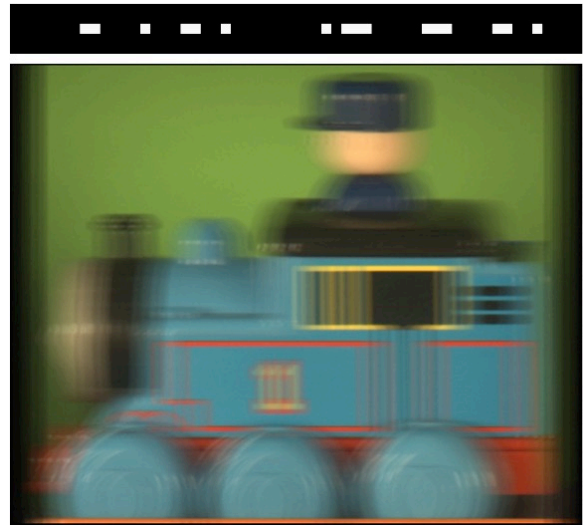
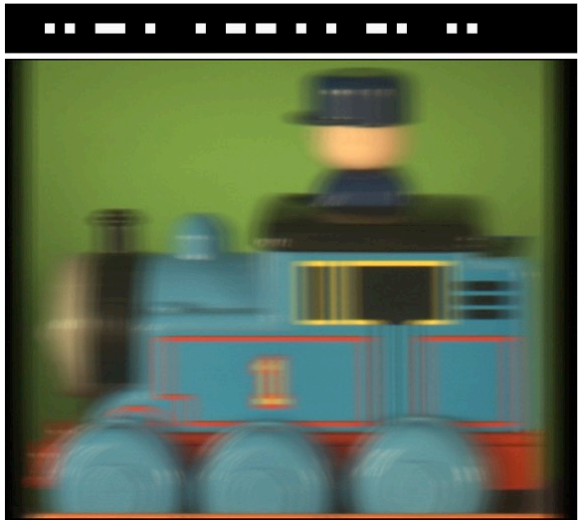 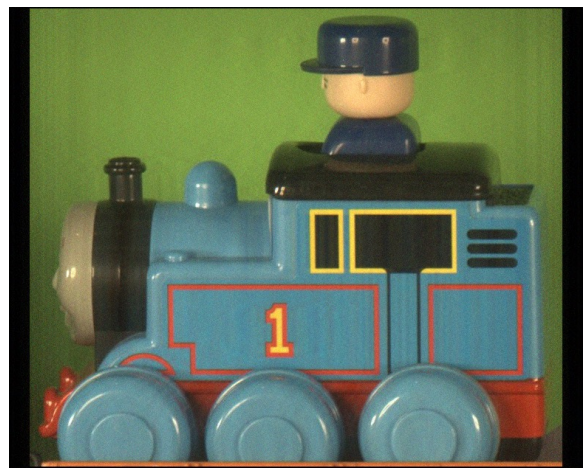

**Figure 17 - Triplet of interleaved exposure images and resting average of deblurred images.**

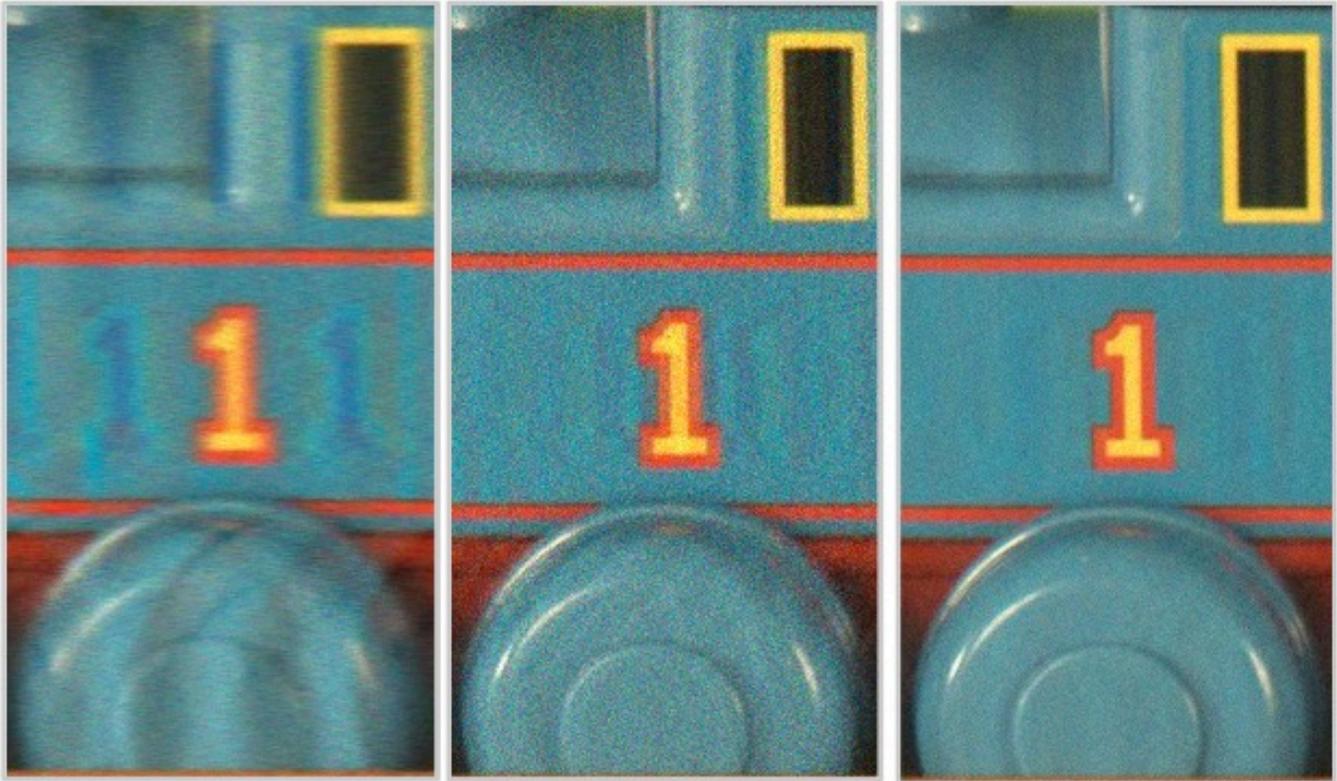

Figure 18 - Detail: Flat. Raskar's, Triplet

**Comment: 22-Aug-2015**

**Triplet (or double) exposure can be achieved in controlled environment and mostly white / gray objects (for example black text on white background on a conveyor belt) using Red Blue and Green flickering lights and a conventional color camera.**

# Appendix A - Testing script

```
clear
nf = 0.01; % noise mean and std
org_img=im2double(imread('target.jpg'));

% ============== flat PSF ===================

psf=im2double(imread('flat.pgm'));
psf=psf/(sum(sum(psf)));

dc = 1.0;               % 100% duty cycle
img = org_img * dc ;

blr(:,:,1) = conv2(img(:,:,1),psf);
blr(:,:,2) = conv2(img(:,:,2),psf);
blr(:,:,3) = conv2(img(:,:,3),psf);

% duty cycle dependent noise
noise = nf*dc + nf*dc*randn(size(blr));
blrn = blr + noise;

imwrite(blrn,'blur-flat.png');
flat = deconvlucy(blrn,psf,20);

% ============== Raskar (original) PSF ===================

psf=im2double(imread('raskar.pgm'));
psf=psf/(sum(sum(psf)));

dc = 0.25;              % 50% duty cycle 1/2 shutter attenuation
img = org_img * dc ;

blr(:,:,1) = conv2(img(:,:,1),psf);
blr(:,:,2) = conv2(img(:,:,2),psf);
blr(:,:,3) = conv2(img(:,:,3),psf);

% duty cycle dependent noise
noise = nf*dc + nf*dc*randn(size(blr));
blrn = blr + noise;

imwrite(blrn,'blur-raskar.png');
imwrite(blrn/dc,'blur-raskar-scaled.png');
raskar = deconvlucy(blrn,psf,20);

raskar = raskar / dc ;
```

```matlab
% inverse Raskar PSF

psf=im2double(imread('raskarInv.pgm'));
psf=psf/(sum(sum(psf)));

dc = 0.25;              % 50% duty cycle 1/2   shutter attenuation
img = org_img * dc ;

blr(:,:,1) = conv2(img(:,:,1),psf);
blr(:,:,2) = conv2(img(:,:,2),psf);
blr(:,:,3) = conv2(img(:,:,3),psf);

% duty cycle dependent noise
noise = nf*dc + nf*dc*randn(size(blr));
blrn = blr + noise;

imwrite(blrn,'blur-raskarInv.png');
imwrite(blrn/dc,'blur-raskarInv-scaled.png');
raskarInv = deconvlucy(blrn,psf,20);

raskarInv = raskarInv / dc ;

raskarBoth = (raskar + raskarInv) / 2;

% ============== Triplet PSF ====================

psf=im2double(imread('s1.pgm'));
psf=psf/(sum(sum(psf)));

dc = 0.3077;            % 30.77 duty cycle
img = org_img * dc ;

blr(:,:,1) = conv2(img(:,:,1),psf);
blr(:,:,2) = conv2(img(:,:,2),psf);
blr(:,:,3) = conv2(img(:,:,3),psf);

% duty cycle dependent noise
noise = nf*dc + nf*dc*randn(size(blr));
blrn = blr + noise;

imwrite(blrn,'blur-s1.png');
imwrite(blrn/dc,'blur-s1-scaled.png');
s1 = deconvlucy(blrn,psf,20);

s1 = s1 / dc;

psf=im2double(imread('s2.pgm'));
psf=psf/(sum(sum(psf)));

dc = 0.3462;            % 34.62% duty cycle
img = org_img * dc ;

blr(:,:,1) = conv2(img(:,:,1),psf);
blr(:,:,2) = conv2(img(:,:,2),psf);
blr(:,:,3) = conv2(img(:,:,3),psf);
```

```matlab
% duty cycle dependent noise
noise = nf*dc + nf*dc*randn(size(blr));

blrn = blr + noise;

imwrite(blrn,'blur-s2.png');
imwrite(blrn/dc,'blur-s2-scaled.png');
s2 = deconvlucy(blrn,psf,20);

s2 = s2 / dc;

psf=im2double(imread('s3.pgm'));
psf=psf/(sum(sum(psf)));

dc = 0.3462;              % 34.62% duty cycle
img = org_img * dc ;

blr(:,:,1) = conv2(img(:,:,1),psf);
blr(:,:,2) = conv2(img(:,:,2),psf);
blr(:,:,3) = conv2(img(:,:,3),psf);

% duty cycle dependent noise
noise = nf*dc + nf*dc*randn(size(blr));
blrn = blr + noise;

imwrite(blrn,'blur-s3.png');
imwrite(blrn/dc,'blur-s3-scaled.png');
s3 = deconvlucy(blrn,psf,20);

s3 = s3 / dc;

res = (s1 + s2 + s3) / 3;

%=========== average 3 short 17 psf ==============

psf=im2double(imread('flat17.pgm'));
psf=psf/(sum(sum(psf)));

dc = 1.0/3.0;             % 33.333% duty cycle
img = org_img * dc ;

clear blr;

blr(:,:,1) = conv2(img(:,:,1),psf);
blr(:,:,2) = conv2(img(:,:,2),psf);
blr(:,:,3) = conv2(img(:,:,3),psf);

% duty cycle dependent noise
noise = nf*dc + nf*dc*randn(size(blr));
blrn = blr + noise;

imwrite(blrn,'blur-a1.png');
imwrite(blrn/dc,'blur-a1-scaled.png');
a1 = deconvlucy(blrn,psf,20);

a1 = a1 / dc;
```

```
psf=im2double(imread('flat17.pgm'));
psf=psf/(sum(sum(psf)));

dc = 1.0/3.0;              % 33.333% duty cycle
img = org_img * dc ;

blr(:,:,1) = conv2(img(:,:,1),psf);
blr(:,:,2) = conv2(img(:,:,2),psf);
blr(:,:,3) = conv2(img(:,:,3),psf);

% duty cycle dependent noise
noise = nf*dc + nf*dc*randn(size(blr));
blrn = blr + noise;

imwrite(blrn,'blur-a2.png');
imwrite(blrn/dc,'blur-a2-scaled.png');
a2 = deconvlucy(blrn,psf,20);

a2 = a2 / dc;

psf=im2double(imread('flat17.pgm'));
psf=psf/(sum(sum(psf)));

dc = 1.0/3.0;              % 33.333% duty cycle
img = org_img * dc ;

blr(:,:,1) = conv2(img(:,:,1),psf);
blr(:,:,2) = conv2(img(:,:,2),psf);
blr(:,:,3) = conv2(img(:,:,3),psf);

% duty cycle dependent noise
noise = nf*dc + nf*dc*randn(size(blr));
blrn = blr + noise;

imwrite(blrn,'blur-a3.png');
imwrite(blrn/dc,'blur-a3-scaled.png');
a3 = deconvlucy(blrn,psf,20);

a3 = a3 / dc;

resA = (a1 + a2 + a3) / 3;
```

```matlab
%============== Finalize =========================

figure, imshow([flat raskar raskarBoth res resA]);
title('flat raskar res resA');
drawnow

imwrite(flat,'res-flat.png');
imwrite(raskar,'res-raskar.png');
imwrite(raskarInv,'res-raskarInv.png');
imwrite(raskarBoth,'res-raskarBoth.png');

imwrite(s1,'res-s1.png');
imwrite(s2,'res-s2.png');
imwrite(s3,'res-s3.png');
imwrite(res,'res-s.png');

imwrite(a1,'res-a1.png');
imwrite(a2,'res-a2.png');
imwrite(a3,'res-a3.png');
imwrite(resA,'res-a.png');
```

## Appendix B - Optimization code

```c
//  Created by Moshe Benezra on 8/21/12.
//

#include <stdio.h>
#include <stdlib.h>
#include <math.h>
#include <memory.h>
#include <time.h>

#define B   3           // bits
#define W   52          // gene length
#define N   64          // first power of 2 > W
#define P   1000        // population size

#define SQR(x) ((x)*(x))

void fft(
         unsigned  NumSamples,
         int       InverseTransform,
         float     *RealIn,
         float     *ImagIn,
         float     *RealOut,
         float     *ImagOut );

static int e[5] = {1,2,4,8,16}; // valid element 1 bit on of n

static struct gene {
    unsigned  v[W];
    float score;
} pop[P], next[P];

void generate()
{
    for (int i=0; i<P; i++){
        for (int j=0; j<W; j++) {
            
            int k = random() % B;
            pop[i].v[j]=e[k];
            
        }
        pop[i].score = i;
    }
}
```

```c
void score()
{
    float inRe[N],  inIm[N];
    float outRe[N], outIm[N];

    float min[B];
    float max, max1, max2;

    for (int i=0; i<P; i++) {
        pop[i].score = -1e6;

        for (int s=0; s<B; s++) {
            memset(inRe, '\0', sizeof(inRe));
            memset(inIm, '\0', sizeof(inIm));

            for (int j=0; j<W; j++){
                inRe[j] = (pop[i].v[j] >> s) & 0x1;
            }

            fft(N,0,inRe, inIm, outRe, outIm);

            min[s] = 1e6;

            for (int j=0; j<=N/2; j++) {
                float t = 20.0*log10(sqrtf(SQR(outRe[j])+SQR(outIm[j])) /
                          (N/2));
                if (min[s]>t) min[s] = t;
            }
        }

        max = -1e6;
        for (int j=0; j<B; j++){
            if (max < min[j]) max = min[j];
        }

//      pop[i].score = max;
//      pop[i].score = (min[0]+min[1]+min[2])/3.0;

        max1 = (min[0] > min[1] ? min[0] : min[1]);
        max2 = (min[0] > min[2] ? min[0] : min[2]);

        max = (max1 + max2) / 2;
        pop[i].score = max;
    }
}
```

```c
void sort()
{
    gene tmp;

    for (int i=0; i<P-1; i++) {
        for (int j=i+1; j<P; j++) {
            if (pop[i].score < pop[j].score){
                tmp = pop[i];
                pop[i]= pop[j];
                pop[j]=tmp;
            }
        }
    }

}

void report(int g)
{
    printf("Generation %04d: Score: %06.4f ", g, pop[0].score);
    for (int i=0; i<W; i++) printf("%0x",pop[0].v[i]);
    putchar('\n');
}

void select()
{
    // ellite
     for (int i=0; i<3 ; i++) {
        next[i] = pop[i];
    }

    float sum = 0;

    for (int i=0; i<P; i++)
        sum += 1000 + pop[i].score;   // (make sum positive)

    for (int i=3; i<P; i++) {
        int r = random() % (int) sum;

        float tsum=0;
        for (int j=0; j<P; j++) {
            tsum += 1000 + pop[i].score;
            if (tsum > r) {
                next[i] = pop[j];
                break;
            }
        }
    }
    memcpy(next,pop,sizeof(next));
}
void cross()
```

```c
{
    for (int i=3; i<P; i++) {
        if ((random()%1000) < 950) {
            int j,k;
            while ((j= random() % P) < 3);

            k = random() % W;

            for (int l=k; l<W; l++){
                unsigned t =  pop[i].v[l];
                pop[i].v[l] = pop[j].v[l];
                pop[j].v[l] = t;
            }
        }
    }
}

void mutate()
{
    for (int i=3; i<P; i++) {
        for (int j=0; j<W; j++) {
            if ((random()%1000) < 15) {
                int k = random() % B;
                pop[i].v[j] = e[k];
            }
        }
    }
}

int main()
{
    srandom(time(NULL));

    generate(); // generate population

    for (int i=0; i< 10000000; i++) {

        score();        // set fitness number
        sort();         // rank current population
        report(i);      // report best
        select();       // copy and clone top
        cross();        // crossover
        mutate();       // mutate some of all new population

    }

    return 0;
}
```